# Irregular curvature at focal adhesions modulates Piezo1 activity and low-frequency ultrasound–induced apoptosis in cancer cells

Ivana Pajic-Lijakovic [1,4], Milan Milivojevic [1], Boris Martinac [2,3], and Peter V. E. McClintock [4]

[1] University of Belgrade, Faculty of Technology and Metallurgy, Department of Chemical Engineering, Belgrade, Serbia

[2] Mechanosensory Biophysics Laboratory, Victor Chang Cardiac Research Institute, Sydney, Australia

[3] St Vincent's Clinical School, Faculty of Medicine, University of New South Wales, Sydney, New South Wales, Australia

[4] Department of Physics, Lancaster University, Lancaster LA1 4YB, UK

Correspondence to: Ivana Pajic-Lijakovic, iva@tmf.bg.ac.rs and,

Peter V. E. McClintock, p.v.e.mcclintock@lancaster.ac.uk

**Abstract**

Low-frequency, low intensity ultrasound (LIUS) has emerged as a promising physical modality capable of inducing selective apoptosis of cancer cells, while sparing healthy epithelial cells and fibroblasts. Hitherto, the mechanism underlying this selectivity has been unclear, but we now propose and develop a theoretical framework linking the distinct mechanical behaviours of cancer versus healthy cells to their differential responses to LIUS. We point out that cancer cells exhibit inhomogeneous ventral stress-fiber networks, which can produce irregular focal adhesion geometry and inward membrane curvature near focal adhesions under low-intensity ultrasound (LIUS). These curvature irregularities can favor loose packing of Piezo1 channels, thereby preserving their activity. In contrast, healthy epithelial cells and fibroblasts display more homogeneous cytoskeletal organization, which can result in more regular curvature profiles adjacent to focal adhesions. This leads to curvature-driven cholesterol redistribution, resulting in altered spatial organization of Piezo1 clusters and reduced coordinated channel activity and allowing cells to remain in their active, proliferative state when exposed to LIUS. Based on theoretical modeling and previous experimental findings, we propose that differences in cytoskeletal organization and membrane curvature can contribute to distinct Piezo1 activation patterns between healthy and cancerous cells. Our analysis identifies curvature-mediated Piezo1 redistribution as a potential physical basis for LIUS selectivity and provides a mechanistic foundation for designing ultrasound-based therapies to exploit the intrinsic cytoskeletal vulnerabilities of cancer cells.

**Glossary of terms**

**Anisotropy:** refers to the characteristic of a system in which its physical properties differ based on the direction of measurement. This indicates that a system may display varying attributes, such as strength or stiffness, along distinct axes.

**Cadherins**: transmembrane glycoproteins characterized by an extracellular domain that facilitates cell-cell adhesion through either homophilic or heterophilic interactions, as well as an intracellular domain that regulates signaling cascades associated with numerous cellular processes, such as polarity and gene expression, among others.

**Cell mechanosensing:** is the ability of cells to detect and respond to mechanical cues—such as stiffness, tension, compression, or shear—through specialized molecular structures like integrins, the cytoskeleton, and mechanosensitive ion channels.

**Epithelial cells**: are cells that exhibit cuboidal shape, limited mobility, apical-basal polarity, and strong E-cadherin-mediated cell-cell adhesions. These cells undergo collective migration during biological processes such as morphogenesis and wound healing.

**Focal adhesion**: a multi-protein complex that mechanically links the cell's cytoskeleton to the extracellular matrix through integrins, enabling the cell to sense forces, generate traction, and regulate migration.

**Integrin**: a transmembrane protein that constitutes part of a focal adhesion complex.

**Integrin tilting**: A change in the orientation of integrin receptors relative to the plasma membrane and underlying actin cytoskeleton, such that the integrin–extracellular matrix linkage becomes obliquely aligned rather than normal to the membrane or substrate

**Mechanical stress**: a physical quantity that describes the magnitude and direction of the force per unit area causing a system deformation.

**Membrane surface tension**: an indicator of the membrane's cohesiveness, which is affected by the dilational viscoelastic properties of the lipid bilayer and the actin cortex, in addition to the local curvature that results from the bending of the lipid bilayer.

**Membrane inward curvature around focal adhesions**: a localized bending of the plasma membrane toward the cell interior at sites of focal adhesions, caused by the mechanical coupling between integrins, the actin cytoskeleton, and the extracellular matrix. This inward curvature concentrates forces at adhesion points, influences focal adhesion maturation, and can modulate signaling and cytoskeletal organization in the surrounding ventral membrane.

**Mesenchymal-like cancer cells**: tumour cells that exhibit elongated shapes, increased migratory ability, front-rear cell polarity, and weak N-cadherin-mediated cell-cell adhesion.

**Piezo1 channel:** a large mechanosensitive homotrimeric membrane protein with a triskelion-like structure comprising extracellular blades, a cap, intracellular anchors, beams, and a pore.

**Strain**: the deformation of a system in response to mechanical stress.

**Tilting of the ventral stress fiber network:** a collective reorientation of ventral actin stress fibers, anchored to focal adhesions at the basal (ventral) surface of the cell, such that the fiber network becomes inclined or biased relative to the cell's main axes.

**Ventral stress fibers:** consist of bundles formed by actin filaments and non-muscle myosin II filaments. These fibers are secured at both ends to focal adhesions, thereby linking adjacent focal adhesions across the ventral cytoskeletal surface. Focal adhesions function as mechanical junctions that integrate the contractile stress fibers into a cohesive, load-bearing network.

**Viscoelasticity**: A time-dependent response of cell membrane under fluctuations, which includes energy storage and dissipation during its structural changes. Mechanism of energy storage and dissipation is closely connected with the stress and strain relaxation phenomena and can be described in the form of proper constitutive model.

## 1.Introduction

The efficacy of cancer therapies like radiation and chemotherapy is frequently restricted. Their key limitations are: (i) the nonselective harm inflicted on both epithelial and cancer cells, and (ii) the increased transition of cancer cells from a proliferative to a quiescent state. Quiescent cancer cells, which are described as 'dormant', non-dividing, and inactive in the cell cycle, are viewed as a crucial element in the ineffectiveness of cancer pharmacotherapies [1]. Ionizing radiation can enhance the migration and invasion of cancer cells through complex interactions within the microenvironment, including changes in cell–cell junctions, modifications to extracellular matrix connections, the secretion of proteases, and triggering of the epithelial–mesenchymal transition [2,3].

In recent years, biophysical approaches have emerged as promising complementary strategies to overcome these limitations by exploiting mechanosensitive pathways that are differentially activated in cancer versus healthy epithelial cells [4-7]. Low-frequency and low intensity ultrasound (LIUS) represents one such approach. Unlike cytotoxic therapies that target proliferation, LIUS can directly perturb the mechanical state of the cytoskeleton, focal adhesions (FAs), and associated mechanotransducers, thereby influencing the fate of cancer cells through physical rather than biochemical selectivity [6, 8-11]. A growing body of evidence suggests that cancer cells exhibit increased intracellular heterogeneity across multiple scales, including nanoscale disorder signatures derived from optical refractive index fluctuations [12,13], together with altered actomyosin contractility [14] and spatially heterogeneous focal adhesion mechanics [15]. These features are regulated by multiple signaling pathways, including intracellular calcium dynamics [16], which contribute to the temporal modulation of cytoskeletal tension and FA remodelling. While these mechanical measurements do not directly resolve stress fiber structures, they provide indirect constraints on the organization of the underlying contractile cytoskeleton, as stress fibers represent one of several actomyosin-based structures contributing to the generation of traction forces in adherent cells. Together, this creates local variations in accumulated mechanical stress [17] resulting in mechanical irregularities, which sensitize cancer cells—more so than healthy epithelial cells—to ultrasound-induced mechanical stresses [4,6].

Importantly, LIUS has been shown to induce apoptosis preferentially in malignant cells by activating mechanosensitive ion channels such as Piezo1 [6], whose gating depends on: (i) the distribution of Piezo1 channels [5], (ii) mechanical stress [17], (iii) membrane curvature [18] and (iv) concentration of cholesterol within the lipid bilayer [19]. Interestingly, LIUS perturbs the homogeneous distribution of Piezo1 molecules along the membranes of mesenchymal-like cancer cells leading to the clustering of Piezo1 molecules around FAs [6]. This clustering is a necessary but not sufficient condition for the apoptosis of cancer cells, corresponding to the fact that Piezo1 molecules are grouped around FAs in epithelial cells and fibroblasts both with and without LIUS. These cell types maintain their active, proliferative state under LIUS [5,6]. This distinctive behaviour of cancer cells in comparison with epithelial cells and fibroblasts is closely connected with the activity of Piezo1 channels. While Piezo1 channels grouped around FAs in cancer cells are overactive [6], they are less active or even inactive in epithelial cells [20-22]. Thus, LIUS may bypass the problem of cancer cell quiescence by engaging mechanical pathways that remain active regardless of cell cycle state, thus offering a route to eliminate dormant or slow-cycling cancer cell subpopulations.

Furthermore, recent studies indicate that the mechanical microenvironment of tumors—including matrix stiffness, viscoelasticity, and anisotropic stress transmission—strongly modulates the response to physical perturbations. Mesenchymal-like cancer cells, which typically show enhanced invasiveness and traction force heterogeneity, may be especially susceptible to LIUS-induced destabilization of focal adhesion–cytoskeleton coupling. By perturbing the resistive force caused primarily by remodeling the ventral cytoskeleton and by driving irregular curvature formation along FAs, LIUS can disrupt mechanotransductive signaling, reduce migratory capacity, and promote apoptotic pathways [6,23]. This mechanobiological mode of action positions LIUS as a promising adjunct to conventional therapies by targeting the mechanical vulnerabilities intrinsic to cancer cell structure.

The main objective of this theoretical study is to identify and discuss the physical mechanisms that underlie the heightened vulnerability of cancer cells to LIUS, in contrast to healthy epithelial cells and fibroblasts, which largely preserve their active and proliferative states. We particularly emphasize how curvature geometry governs the spatial distribution of Piezo1 channels around FAs, thereby modulating their mechanosensitive activity.

## 2. Characteristics of LIUS

Continual or pulsatile LIUS utilized across different tissues, is characterized by an intensity ranging from 10-120 mW·cm$^{-2}$ and frequency in the range of 20 kHz to 1 MHz. The wavelength of LIUS is 1.5 mm at 1 MHz and longer for lower frequencies. The pulse duration for pulsatile LIUS is 100-200 $\mu s$, while the duty cycle is typically 50% and even less, implying that the ultrasound is "on" only for part of each cycle [6,8-11]. The exposure duration varies from a few minutes to a few hours [4,6]. LIUS induces the generation of longitudinal pressure waves. The corresponding acoustic pressure amplitude can be expressed as: $\Delta p = \sqrt{2\rho c I}$ (where $I$ is the intensity of ultrasound, $c$ is the speed of ultrasound in water-like media $c \approx 1500\ \frac{\mathrm{m}}{\mathrm{s}}$, $\rho$ is the density of multicellular systems $\rho \approx 1000\ \frac{kg}{m^3}$). The pressure amplitude is equal to: 550 Pa for an ultrasound intensity of 10 mW·cm$^{-2}$ and 1.7 kPa for an intensity of 100 mW·cm$^{-2}$. Acoustic pressure oscillations induce successive compression and extension of tissue and shear stress caused by fluid flow around cells. The intensity and frequency of LIUS influence the resultant displacement amplitude: $\Delta r_{LIUS} = \sqrt{\frac{2I}{\rho c \omega^2}}$. Consequently, for an intensity of 100 mW·cm$^{-2}$ and frequency of 33 kHz, the displacement amplitude is 180 nm, while for a frequency of 120 kHz, the displacement amplitude is 50 nm.

The biological response to ultrasound is strongly regime-dependent. While the present work focuses on low-intensity ultrasound (LIUS), where mechanical perturbations remain within the non-destructive regime, it is important to note that higher intensities or continuous exposure can lead to qualitatively different outcomes, including cytoskeletal disruption and FA disassembly. In contrast, the LIUS regime considered here primarily induces reversible mechanical effects such as modulation of actomyosin contractility and adhesion dynamics. A key aspect of the mechanism is the translation of macroscale acoustic fields into localized subcellular responses. Although the acoustic wavelength is on the millimeter scale, the induced displacement amplitudes (on the order of tens to hundreds of nanometers) are comparable to the dimensions of FAs and cytoskeletal structures. These displacements are transmitted through integrin-mediated adhesions to the actomyosin network, where they are amplified by intracellular force generation and mechanical heterogeneity. As a result, relatively small global deformations can produce localized stress imbalances and nanoscale membrane curvature variations at focal adhesions, thereby influencing Piezo1 localization and activation.

Pulsatile and low-frequency LIUS is typically operated in a non-thermal, non-cavitation regime by appropriate control of the acoustic intensity and duty cycle. Here is the reasoning:

- Cavitation (formation and collapse of microbubbles) generally requires higher acoustic pressures or continuous wave exposure. In LIUS, the intensity is usually below 100 mW·cm$^{-2}$, which is too low to induce stable or inertial cavitation in biological tissues.
- Temperature rise is also minimal because the ultrasound is pulsed rather than continuous — allowing heat to dissipate between pulses. In most cell culture and tissue studies, the temperature increase is <0.1–0.2 °C, which can be considered negligible.
- As a result, mechanical effects — such as radiation force, acoustic streaming, and cyclic deformation of the cell membrane and cytoskeleton — dominate, rather than thermal effects.

The response of multicellular systems to LIUS occurs on both: (i) a short timescale from microseconds to milliseconds and (ii) a long timescale from minutes to hours. As already noted, the period of the ultrasound oscillations corresponds to microseconds, while the time between two pulses corresponds to milliseconds. The exposure time is from a few minutes to a few hours. The timescale from microseconds to milliseconds corresponds to protein folding, while a timescale of seconds corresponds to inter-filament interactions and their alignment. The inactivation time of Piezo1 channels is typically on the order of a few tens of milliseconds in heterologous expression systems, such as HEK293T or N2A cells, as reported in patch-clamp and whole-cell recordings [24,25]. However, in native cells, inactivation can be dramatically slower, depending on factors such as binding partners, membrane composition, curvature, and cytoskeletal interactions [26]. Heterologous systems allow precise experimental control but may not fully capture the kinetics observed in physiological environments. Consequently, lower frequencies LIUS $< 40\,\mathrm{kHz}$ are more efficient in the activation of Piezo1 channels [6]. The timescale of seconds corresponds to the bending relaxation of the lipid bilayer [27]. The timescale of minutes corresponds to the remodeling of adhesion complexes [28], and rearrangement of cytoskeleton domains, as well as to the cell shape relaxation time [29] while the time scale of hours corresponds to collective cell migration [17,30,31]. The cell repolarization time, caused by contact inhibition of locomotion, corresponds to times ranging from a few tens of minutes to hours [31]. Post-translational modification of membrane proteins, such as phosphorylation and glycosylation, may only require a few minutes, whereas synthesis of proteins and their transport can take tens of minutes [32]. Upon further examination, we will explore the responses of different cell types to LIUS, drawing on findings from the existing literature.

## 3. Cell response to LIUS

The collective migration and proliferation of the majority of mesenchymal, cancer cells are reduced by LIUS, whereas healthy epithelial cells are stimulated by it. Calcium-induced apoptosis of cancer, cells is primarily caused by lower frequencies $< 120\,\mathrm{Hz}$ [6,8-11]. Higher frequencies reduce conformational changes of mechanosensitive Piezo1 channels. The main characteristics of LIUS and cellular systems are shown in **Table 1**:

**Table 1**. A summary of characteristics of cellular systems and LIUS as well as the outcomes

| Cell type | frequency | LIUS parameters | outcome | References |
|---|---|---|---|---|
| Tumour cell lines (MDA-MB-231, A375p, HT1080) vs normal epithelial (MCF10A, HFF) | 33 kHz;<br>120 kHz | pulsatile<br>39 mW·cm$^{-2}$<br>50% duty cycle<br>2 h exposure time | At 33 kHz: significant calcium induced tumour cell apoptosis;<br>at 120 kHz: no increase in apoptosis.<br>Healthy epithelial cells were unaffected. | Tijore et al. [6] |
| PANC-1 cells (i.e., epithelioid carcinoma cell line derived from the human pancreas) | 30-40 kHz<br>1 MHz | continuous<br><100 mW·cm$^{-2}$<br>10 min to 30 min | inhibition of collective cell migration with effects seen for 48–72 h after irradiation | González et al. [11] |
| Breast cancer mesenchymal-like MDA-MB-231 cells | 3 MHz | Pulsatile<br>50 mW·cm$^{-2}$<br>8 pulses, 10% duty cycle | reduced migration and invasion of MDA-MB-231 cells | Calero et al. [33] |
| Breast epithelial MCF10A cells<br>Breast cancer epithelial-like | From 400 kHz to 650 kHz | 50, 100, and 500 mW·cm$^{-2}$ | Epithelial MCF10A cells retain their active proliferative state, while the viability of cancer MCF7 cells is significantly reduced for the | Ivone et al. [23] |

| MCF7 cells | | | frequencies higher than 550 kHz. An increase in frequency from 400 kHz to 480 kHz causes about 70% increase in the Young modulus of MCF7 cells. Further increase in frequency causes softening of MCF7 cells, quantified by a decrease in the Young's modulus. | |
|---|---|---|---|---|
| keratinocytes (HaCaT) | 500 KHz | Pulsatile, with the number of cycles from 100 to 500,<br>1 min exposure time | Migration and proliferation of cells are enhanced | Leng et al. [34] |
| Breast cancer mesenchymal-like MDA-MB-231<br>Breast epithelial MCF10A cells<br>A375p (metastatic melanoma) | 33 kHz | Pulsatile<br>7 mW·cm$^{-2}$<br>50% duty cycle<br>2 h exposure time | Calcium-induced apoptosis of cancer cells.<br>Healthy epithelial cells were unaffected. | Singh et al. [4] |

Tijore et al. [6] revealed that pulsatile LIUS (frequency of 33 kHz, for 2 h at a low intensity of 39 mW·cm$^{-2}$; 50% duty cycle) triggers intensive apoptosis of cancer cells such as MDA-MB-231, A375p, HT1080. Singh et al. [4] applied LIUS of similar specification, though at a lower intensity of 39 mW·cm$^{-2}$, and obtained a similar result. The apoptosis of cancer cells is induced through a calpain-dependent mitochondrial pathway that relies upon calcium entry through the mechanosensitive Piezo1 channels, whereas this pathway is not activated in breast epithelial MCF10A cells and fibroblasts retained their intact state.

Cancer cells are more resistant at higher frequencies of LIUS [6,8-11]. Pulsatile LIUS (at 3 MHz, for 8 pulses, 10% duty cycle, with intensity of 50 mW·cm$^{-2}$) reduced migration and invasion of MDA-MB-231 cells [33]. Migration and proliferation of keratinocytes (HaCaT) are enhanced by pulsatile LIUS at 500 KHz with the number of cycles ranging from 100 to 500 and an exposure time 1 min [34]. Collective migration of PANC-1 cells (i.e., an epithelioid carcinoma cell line derived from the human pancreas) during the wound healing process under continuous exposure to LIUS was considered by González et al. [11]. The cell monolayer's exposure to LIUS (frequencies near 1 MHz, low intensities <100 mW·cm$^{-2}$) over 10 min to 30 min induced long-term inhibition of collective cell migration, with effects being seen for 48–72 h after exposure. Ivone et al. pointed out that the viability of breast cancer MCF7 cells is significantly reduced for frequencies higher than 550 kHz and LIUS intensities in the range of 50-500 mW cm$^{-2}$ [23].

To better understand the selective low-frequency ultrasound-induced apoptosis of cancer cells, despite epithelial cells and fibroblasts remaining intact or even exhibiting enhanced growth and migration, it is necessary to identify the physical mechanisms linking LIUS-induced mechanical perturbation to Piezo1 recruitment and activation. Within our framework, LIUS redistributes mechanical stresses through the ventral cytoskeleton and FA network, thereby modulating local viscoelastic resistance and membrane curvature in focal adhesion–associated regions. Because Piezo1 activity is sensitive to membrane mechanics and curvature-dependent mechanical cues, these LIUS-induced structural changes may differentially regulate Piezo1 recruitment and activation in epithelial and mesenchymal-like cancer cells. The following section therefore discusses the principal physical factors governing Piezo1 behaviour under LIUS.

## 4. Physical factors that influence activity of Piezo1 channels under LIUS in epithelial and mesenchymal-like cancer cells

A comprehensive understanding of the variations in the distribution of Piezo1 channels across the membrane – along with their activation in fibroblasts, epithelial cells and mesenchymal-like cancer cells both with and without LIUS – requires a focus on the underlying causes of this behaviour. These encompass the interplay between: (i) the distribution of Piezo1 molecules, (ii) the size, homogeneity, and stability of FAs, (iii) the dilatational viscoelasticity of the membrane and bending of the lipid bilayer, (iv) the structural homogeneity and viscoelasticity of the ventral cytoskeleton, and (v) the stiffness homogeneity and viscoelasticity of substrate matrices.

### 4.1 The distribution of Piezo1 molecules

Piezo1 channels are grouped around FAs in active, contractile epithelial cells, whereas their distribution appears more uniform in mesenchymal-like cancer, cells [5,21]. Nonetheless, pulsatile low-frequency LIUS leads to changes in the distribution of Piezo1 molecules in cancer cells. In the presence of LIUS these molecules become grouped near FAs [6]. It is well known that Piezo1 channels tend to group at inward curved regions of the membrane [35,36]. Curvatures can be generated along FAs when the resistance forces are lower than driving forces. Detailed descriptions of driving and resistive forces will be given in **Section 6**. The size and geometry of the curvature depend on: (i) the size of an FA, (ii) integrin tilting, (iii) the stiffness, viscoelasticity, and topology of substrate matrices, (iv) the local inflow of calcium, and (v) exposure to LIUS.

Piezo1 recruitment likely emerges from interplay between biochemical interactions and curvature-mediated physical effects, rather than from curvature alone. Piezo1 distribution may be influenced by interactions with adhesion-associated proteins, cytoskeletal linkers, and other binding partners within the focal adhesion complex, which can provide specific anchoring sites and contribute to stabilization of the channel at adhesion regions (e.g., integrin–cytoskeleton networks and adaptor complexes; cite relevant adhesion literature) [37]. In addition, Piezo1 function and nanoscale organization have been shown to depend on the membrane lipid environment and cholesterol domains, implying that local membrane composition and physical context also contribute to its localization and activity [19,38]. Importantly, recruitment alone does not determine channel function. Once localized, Piezo1 activation depends on local membrane mechanics: curvature-induced leaflet asymmetry, reduced bending rigidity, and local tension can facilitate conformational changes required for channel opening.

Piezo1 channels located near FAs are almost inactive during the migration of epithelial Madin–Darby Canine Kidney (MDCK) cells on inhomogeneous viscoelastic PAA substrates [20,22]. The exposure of Piezo1 molecules near FAs in various epithelial cells to LIUS does not trigger their activation. It is in accordance with the fact that epithelial cells will tolerate LIUS by maintaining their physiological level of intracellular calcium [23, 34]. Conversely, Piezo1 channels are over-activated near FAs in cancer MDA-MB-231 and A375p cells, resulting in calcium-induced apoptosis, unlike in epithelial MCF10A cells and fibroblasts [6]. Consequently, the clustering of Piezo1 channels near FAs in cancer cells exposed to LIUS is a necessary but insufficient condition for the onset of cell apoptosis. The activation of Piezo1 molecules depends on their distribution within curved regions, while this distribution depends on the geometry of the curvatures.

LIUS can cause inhomogeneous tilting of integrin molecules within a Piezo1 channel cluster [39]. The tilted integrin and the inward bending of the bilayer modify the orientation of talin and vinculin, which are linked to the cytoskeleton, resulting in a "pushing" effect on neighboring actin filaments. This effect impacts the local rearrangement of ventral stress fibers [40]. Consistent with the observation that the actin cytoskeleton experiences pre-stress due to myosin II contractility, even minor alterations in attachment geometry can induce local curvature in ventral stress fibers [41]. This process recruits and aligns actin filaments beneath the adhesion, thereby shaping the cytoskeletal curvature around the adhesion rim. Inhomogeneous structure of the ventral stress fiber network influences the shape of the curvature.

However, whether this bending leads to membrane curvature also depends on the structural inhomogeneity and viscoelasticity of the ventral cytoskeleton, a network distinct from the cortex, which can resist curvature formation and thus modulate the resulting membrane shape.

**4.2 The distribution and activity of Piezo1 channels within curved regions along FAs**

Although Piezo1 is generally activated at FAs, highly curved areas of the cell membrane surrounding the adhesions may cause reduced channel activity due to curvature-mismatch or local regulation mechanisms [5,35]. This is because FAs are not static structures but rather specialized, transient junctions. They balance the strong cell attachment to a substrate with the cellular motility. Thus, despite that FAs are sites of high Piezo1 activity, the pronounced membrane curvature in surrounding regions, such as filopodia or tethers, can inhibit or regulate Piezo1 function. Reduced activity of Piezo1 channels in curved membrane regions surrounding FAs may arise through several mechanisms: (i) geometrical constraints caused by the channel clustering [42] that limit the space required for flattening of the dome-shaped channel structure [43], (ii) alterations of long-lasting interactions of the channel with specific lipid types [38], (iii) curvature-driven redistribution of cholesterol, which alters the local membrane organization surrounding Piezo1 channels [19,38,44], and (iv) curvature-induced conformational changes of Piezo1 channels that directly modulate their gating behaviour [45]. Beaven et al. [46] pointed out that cholesterol can favor concave curvature. Concurrently, membrane cholesterol helps maintain a mechanically favorable bilayer environment, thereby supporting channel activity [19,37,38]. These studies demonstrated that Piezo1 channels can function as independent mechanotransducers; however, they did not consider their organization into clusters forming pit-shaped invaginations. In the plasma membrane Piezo1 can still function independently. In addition, these clusters can be regulated by the cytoskeleton [42]. Furthermore, Jiang et al. [47] pointed out that the curved, dome-like architecture of Piezo1 channels induces a local membrane footprint, and that membrane curvature directly modulates channel conformation, thereby influencing gating behaviour. In MscL channels, activation of one channel can influence the gating of nearby channels under sufficient membrane tension. Notably, in the MscL case, among 11 channels situated within an area of $1.6\ \mu m^2$, only 5 channels were concurrently activated [45]. For Piezo1, channels appear to function largely independently [37], but membrane curvature, cholesterol, and local density could provide mechanisms for indirect coordination. Consequently, the activity of Piezo1 channels depends strongly on the curvature geometry, which influences the size, shape, and topology of the curvature around FAs.

The purpose of our numerical estimates of packing density of Piezo1 molecules and radius of curvature around FAs is threefold. Firstly, they provide order-of-magnitude validation that the proposed mechanism operates within physically and biologically realistic regimes. Secondly, they establish geometric compatibility between focal adhesion dimensions, membrane curvature scales, and the intrinsic size and curvature sensitivity of Piezo1 channels. Thirdly, they offer a quantitative framework for interpreting how local curvature and channel density may influence Piezo1 clustering and activation, thereby linking the theoretical model to experimentally accessible parameters and guiding future measurements. Yao et al. [5] pointed out that the concentration of Piezo1 channels in epithelial cells is 2-6 times higher near FAs than in regions far from FAs. The average concentration of Piezo1 channels is $\langle c_p \rangle \sim 2\ \mu\mathrm{m}^{-2}$. Consequently, the concentration of Piezo1 channels near an FA is $c_p^{FA} = 4 - 12\ \mu\mathrm{m}^{-2}$. The number of Piezo1 channels per individual FA can be calculated as $N_p^{FA} = c_p^{FA} A_{eff}$ (where $N_p^{FA}$ is the number of Piezo1 channels near FA, and $A_{eff}$ is the effective surface area of the membrane around FA). As a first approximation, we can introduce the effective surface area $A_{eff}$ around an FA as being equal to the surface area of a single FA, i.e. $A_{eff} = Lw$ (where $L$ is the length and $w$ is the width of an FA). For an FA length $\sim 5\ \mu\mathrm{m}$ and a width of $w = 300\ \mathrm{nm}$ [48], the number of Piezo1 channels is $N_p^{FA} \sim 6 - 18$. Piezo1 forms a homotrimeric dome-shaped channel in the plasma membrane. The effective radius of its footprint is $\sim 25 - 50\ \mathrm{nm}$ [43]. The surface area around an FA covered by Piezo1 channels is 3-9% of the effective surface area $A_{eff}$. In accordance with the fact that $L \gg w$, cylindrical shaped curvature can be expected. It occurs mainly across the short axis, i.e., $w \approx R\theta$ (where $R$ is the radius of curvature and $\theta$ is the central angle of the circular arc). The curvature height $h = R\left(1 - \cos\frac{\theta}{2}\right)$. For small deformation $h \ll w$, $h \approx R\frac{\theta^2}{8}$, while $\vartheta = \frac{w}{R}$. Consequently, the radius of curvature $R \approx \frac{w^2}{8h}$, so that $R = 100 - 1000\ \mathrm{nm}$ for $h = 10 - 100\ \mathrm{nm}$. These values overlap with the intrinsic curvature scale of Piezo1 (~50–100 nm) inferred from its dome-shaped structure [49] and fall within the regime where membrane mechanics, including curvature and tension, are known to influence mechanosensitive channel gating and clustering [19].

Overall, two types of curvature can be generated around FAs in cell membranes: regular and irregular as shown in **Figure 1**:

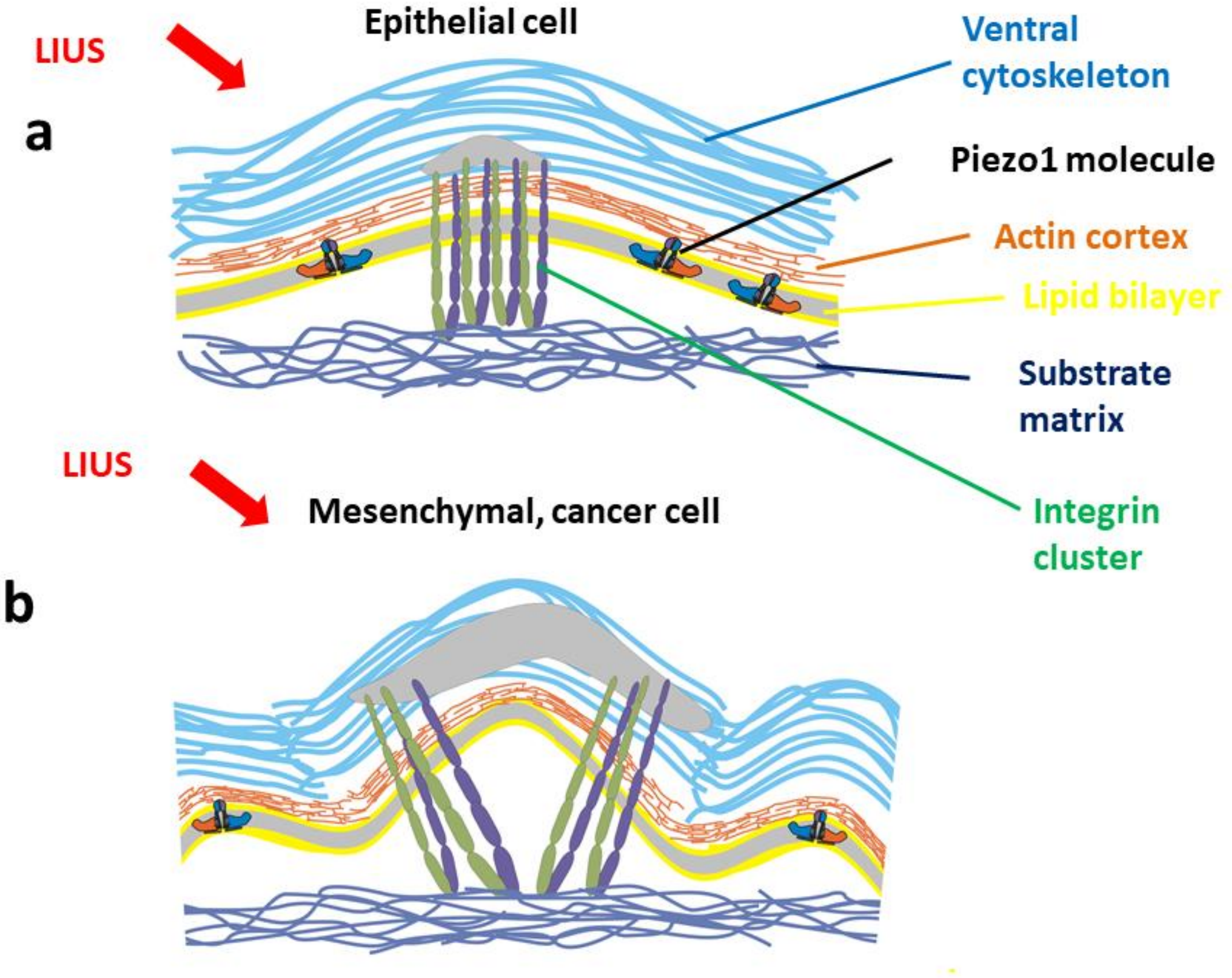

**Figure 1**. (a) Regular curvature, and (b) irregular curvature along a focal adhesion (FA). While epithelial cells generate relatively regular membrane curvatures along FAs, mesenchymal-like cancer cells can produce more irregular curvatures in these regions due to the tilting of integrin clusters and the ventral stress fiber network. Piezo1 channels exhibit a closed packing arrangement within regular curvatures. Conversely, the packing of Piezo1 in irregular curvatures is characterized by looser configurations.

Regular curvatures can be described by one or two approximately constant radii, corresponding to smoothly varying membrane deformations over length scales much larger than individual membrane proteins. In contrast, irregular curvature consists of multiple localized sub-curvature regions, forming a heterogeneous curvature field with spatial variations on length scales of tens to hundreds of nanometers. Notably, these length scales are comparable to or moderately larger than the footprint and intrinsic curvature sensitivity of Piezo1 (~10–100 nm), implying that individual channels may experience distinct local curvature environments. The irregular curvature surrounding FAs arises from interplay between the non-uniform tilting of integrin molecules within FAs and the irregular structural arrangement of the ventral stress fiber network. These structural irregularities are attributable to: (i) internal factors, including the type of cells, and (ii) external factors, such as the non-uniform distribution of matrix stiffness and topology, as well as to the influence of external forces like LIUS. Furthermore, it is necessary to discuss the main cellular constituents responsible for the generation of curvature such as: (i) the size and stability of FAs in various cell types and (ii) the main characteristics of the ventral cytoskeleton, which can resist curvature formation to some extent.

### 4.3 The size and stability of FAs in various cell types

While epithelial cells stabilize their migration by using FAs and strong E-cadherin–mediated cell–cell adhesion contacts, mesenchymal-like cancer cells and fibroblasts stabilize their migration mainly by using FAs. Some mesenchymal-like cancer cells form weak N-cadherin mediated cell-cell adhesion contacts and migrate as loosely coordinated cells, while the others, similarly to fibroblasts, migrate as free cells [50]. Consequently, the sizes of FAs are larger in cancer cells and fibroblasts in comparison with epithelial cells, even when they are cultured on the same substrate matrix. Despite the FAs being large in mesenchymal-like cancer cells, they are less stable than those in fibroblasts and epithelial cells [51].

The stability of FAs depends on the intracellular concentration of calcium. Epithelial and cancer cells exhibit distinct mechanisms for the regulation of intracellular calcium concentration. Epithelial cells are capable of sustaining their intracellular calcium levels within a consistent, physiological range. Conversely, the intracellular calcium levels in cancer cells fluctuate at a frequency of 0.01 Hz [52,53]. A higher concentration of calcium activates calpain, a calcium-dependent protease that can cleave talin, vinculin, and various other components of FAs [16,54,55]. Consequently, the oscillations of intracellular calcium concentration in cancer cells make their FAs less stable than those of epithelial cells [16]. Their unstable FAs enable cancer cells to achieve more efficient collective cell migration.

Although epithelial cells generate smaller FAs compared to fibroblasts and cancer cells, these FAs are capable of inducing membrane curvature. In contrast to epithelial cells, cancer cells that establish larger FAs cannot form curvatures along FAs in experiments without LIUS. It seems that other factors play a dominant role in the formation of curvatures along FAs. One of these factors is the ventral cytoskeleton, which resists curvature formation, among other resisting factors such as the lipid bilayer bending and the membrane surface tension described in **Section 6**.

To understand the crucial function of the ventral stress fiber network in the development of curvature along FAs, it is necessary to analyse the primary physical characteristics of the ventral cytoskeleton for these different cell types.

## 5. Ventral stress fiber network: physical characteristics

Ventral stress fibers are bundles of actin filaments and non-muscle myosin II filaments that are interconnected and cross-linked by α-actinin and other actin-binding proteins [56,57]. This structural inter-connection of stress fibers ensures sliding of myosin along the actin fibers. Mature ventral stress fibers consist of 10-30 actin filaments [58]. Ventral stress fibers are anchored at both ends to FAs, and in this way, they connect neighboring FAs across the ventral cell surface [59]. Consequently, in the ventral cytoskeletal network, FAs function as mechanical junctions that connect the contractile stress fiber bundles into an integrated, load-bearing system [57]. The number of matured FAs depends on the cell type and matrix stiffness and is typically a few tens [60]. The stress fibers behave as rigid chains. It is in accordance with fact that the persistence length of filaments is significantly longer than the contour length. Deguchi et al. (2006) considered the tensile properties of single actin-myosin stress fibers isolated from cultured vascular smooth muscle cells [61]. They reported that the average breaking force of a single stress fiber is ~377 nN, while the average Young's modulus is equal to ~1.45 MPa, which is significantly higher than the Young modulus of single cells that corresponds to a few tens of kPa. The viscoelasticity of a single stress fiber has been described by the Kelvin-Voigt model [62]. It means that fiber length can relax under constant stress conditions. The relaxation time is influenced by the elastic modulus and viscosity of the fiber, which are determined by the quantity of actin and myosin filaments grouped together as well as by the extent of crosslinking among them.

Stress fibers are inter-connected and stabilized by actin-binding proteins. The diameter of stress fibers, surface density of stress fibers, and their inter-connectivity can vary along the ventral part of the cytoskeleton. These variations lead to an inhomogeneous distribution of mechanical stress caused by cell tractions, acto-myosin contractions, and the action of LIUS. The ventral stress fiber networks are not composed solely of ventral stress fibers: they can incorporate other stress fiber types, forming a mechanically integrated system [56]. Dorsal stress fibers are typically anchored at one FA and extend toward the cell interior. They can merge with ventral stress fibers, reinforcing the network. Transverse arcs are curved, contractile bundles that are not directly anchored to FAs, but they can interact with ventral stress fibers, contributing to the tension and structural integration of the ventral cytoskeleton. Inhomogeneity of the ventral cytoskeletal network is pronounced in mesenchymal-like cancer cells [12, 13].

### 5.1 Ventral stress fiber networks under LIUS

LIUS induces the remodelling of ventral stress fiber networks [63]. In mesenchymal-like, cancer cells, the stiffness of the cytoskeleton exhibits spatial variation resulting from: (i) localized differences in crosslinking and fiber density, and (ii) inhomogeneous actomyosin contractility [13]. Local stiffness can vary several-fold across the ventral surface; some domains are contractile and rigid, while others are less contractile and softer [64]. LIUS applies oscillatory compression and extension that propagate through multicellular systems. An increase or decrease in stiffness of the ventral cytoskeleton under LIUS depends on several factors:

(i) Network connectivity and steric hindrance: Long aligned filaments produce stronger mechanical coupling and steric constraints under LIUS. These filaments can jam or resist relative sliding leading to stiffening of the cytoskeleton [65]. However, shorter filaments with many orientations more easily rearrange and slide past each other leading to softening of the cytoskeleton.

(ii) Cross-linker dynamics and force-dependent unbinding: If cross-linkers are fast and force-stabilized (by catch bonds or force-inhibited unbinding), mechanical load can transiently increase connectivity and stiffness leading to an increase in the stiffness of the cytoskeleton. If cross-linkers detach under load (slip bonds), softening of the cytoskeleton occurs [66].

(iii) Alignment and hydrodynamic coupling: Aligned bundles transmit energy induced by LIUS more efficiently and can lock into collective translation. Isotropic networks allow local rotation and reorientation that reduce macroscopic resistance.

Because the acoustic field is not perfectly homogeneous on the cellular scale, the stress amplitude and phase vary across regions. In cells with uniform contractility and cytoskeletal stiffness, such oscillations would induce nearly reversible deformation. However, in cells with non-uniform stiffness such as mesenchymal-like cancer cells, each domain exhibits different energy storage and dissipation under strain. The mechanism of mechanical stress accumulation by LIUS can be explained as follows:

- When stiffer domains deform less than softer ones, strain incompatibility arises at their interfaces. This strain incompatibility can induce local tilting along the interfaces [12].
- During cyclic LIUS loading, these local mismatches generate internal stress gradients that do not fully relax after each cycle. This stress accumulation induces additional stiffening of the domains [23], which has a feedback impact on the stability of FAs.

In contrast to the more irregular cytoskeletal organization observed in cancer cells [12], epithelial cells and fibroblasts exhibit distinct self-organized cytoskeletons under isotropic confinement [67,68]. Fibroblasts typically undergo symmetry-breaking to form a homogeneous but anisotropic cytoskeletal organization characterized by aligned ventral stress fibers anchored around mature FAs, often oriented along mechanically preferred directions associated with effective force transmission and substrate sensing [68]. In contrast, epithelial cells tend to maintain a more isotropic actin organization [67]. Schematic presentations of ventral stress fiber networks in epithelial cells and cancer cells are shown in **Figure 2**:

Epithelial cell
a
Focal adhesion
Myosin II
Actin cross-linkers
Ventral stress fiber
Mesenchymal, cancer cells
b

**Figure 2**. Schematic presentation of the ventral stress fiber networks in: (a) epithelial cells and (b) mesenchymal-like cancer cells. Ventral stress fiber networks in epithelial cells are more homogeneous and composed of thinner, more isotropically oriented stress fibers, whereas ventral stress fiber networks in cancer cells exhibit inhomogeneity and more anisotropic orientations. Focal adhesions in mesenchymal-like cancer cells are larger than those in migrating epithelial cells when both are situated on the same substrate matrix.

Sanger et al. compared stress fiber organization in fibroblasts and epithelial cells and found marked structural differences [69]. In fibroblasts, stress fibers are well-developed, thick, long bundles that span large parts of the cell and are strongly associated with FAs. Actin filaments form a fine, dense mesh with short stress fibers (often peripheral) in epithelial cells, but there is no strong global orientation. Stress fibers are stiffer and more contractile in fibroblasts than in epithelial cells. While the ventral stress fiber networks are relatively homogeneous in fibroblasts and epithelial cells, they are inhomogeneous in mesenchymal-like cancer cells [12,13].

Interplay between the inhomogeneous distribution of mechanical stress and the tilting of ventral cytoskeletal domains results in the generation of irregular curvatures near FAs.

### 5.2 Tilting of ventral cytoskeletal domains

The ventral cytoskeleton normally lies almost parallel to the substrate plane (the *x*–*y* plane). A tilt in the *z* direction means that part of this network or domain rotates slightly out of the substrate plane, forming a small angle relative to the substrate. Tilting occurs in migrating cells when there is an imbalance of vertical ($z$-directed) torque acting on the ventral domain. This can happen in several ways: (i) inhomogeneous structure of the ventral cytoskeleton, which leads to an inhomogeneous distribution of mechanical stress, (ii) topological defects of substrate matrices, which cause the generation of asymmetric traction forces, and (iii) action of external forces such as LIUS in the $z$-direction.

Tilting of ventral stress fiber networks during collective cell migration on substrate matrices is closely connected to the cell's ability to provide strong out-of-plane tractions relative to in-plane tractions. In fibroblasts, a radially symmetrical configuration of actin bundles, consisting of α-actinin-rich radial fibers and myosin IIA-rich transverse fibers, spontaneously evolved into a chiral system as a result of the unidirectional tilting of all radial fibers, which was accompanied by a tangential shift in the retrograde movement of transverse fibers [70]. The anisotropy of tractions during cell migration, rather than the magnitude of traction force, is the hallmark of invasive cancer cells such as MDA MB-231 breast carcinoma and A-125 lung carcinoma cells in collagen gels [15].

Within the present framework, tilting of ventral cytoskeletal domains is not only a structural feature but also a key mechanical mechanism that redistributes intracellular stresses. Specifically, tilting modifies the spatial distribution of traction forces and generates local imbalances in the cytoskeletal stress field, captured by the divergence of the stress tensor. These stress inhomogeneities can locally reduce the effective viscoelastic resistance to membrane deformation or, conversely, concentrate stresses in specific regions near focal adhesions. As a result, tilting facilitates the formation of localized and potentially irregular membrane curvatures on submicron length scales. Such curvature variations provide a heterogeneous mechanical environment that can influence the spatial distribution and activation of Piezo1 channels, thereby linking cytoskeletal reorganization to curvature-mediated mechanosensing.

## 6. Theoretical consideration

The mechanical behaviour and spatial organization of Piezo1 channels in the membrane are shaped by both lateral migration and membrane curvature effects. In this section, we first consider the mechanisms driving lateral redistribution of Piezo1 channels, which are governed primarily by the diffusion and gradient of membrane surface tension (the Marangoni effect), modulated further by membrane composition, cholesterol content, and curvature-induced conformational changes. These factors collectively determine the formation, packing, and activity of Piezo1 clusters. We then examine the causes of curvature formation near FAs, analyzing the balance of forces by emphasizing the role of LIUS in the formation of irregular curvature around FAs in cancer cells. Understanding the interplay between Piezo1 migration and curvature provides a framework for predicting channel activity under physiological and pathological conditions [19,21,71,72].

### 6.1 The mechanisms which drive lateral migration of Piezo1 channels

The Piezo1 molecules diffuse laterally along the cell membrane driven by the gradient of their concentration. This results in a uniform distribution of Piezo1 channels, a finding that has been validated in mesenchymal-like cancer cells in the absence of LIUS [5]. The inhomogeneous distribution of Piezo1 channels observed in epithelial cells and fibroblasts in the absence of LIUS [5], alongside mesenchymal-like cancer cells exposed to LIUS [6], suggests the existence of an additional mechanism regulating the migration of Piezo1 channels concurrently with the diffusion mechanism. It originates from the formation of curvature along FAs.

Curvature influences the membrane surface tension. As a measure of the cohesiveness of the membrane, the surface tension, contributes to the reduction of the membrane surface area [17]. The membrane surface tension consists of two components: (i) an in-plane surface tension, which depends on the dilational viscoelasticity of the lipid bilayer and actin cortex and (ii) an out-of-plane surface tension, which depends on the curvature formation. A schematic presentation of the mechanisms that influence the lateral migration of Piezo1 channels is shown in **Figure 3**:

**Lateral movement of Piezo1 molecules**

**Homogeneous distribution**

**Inhomogeneous distribution**

**Gradient of concentration of Piezo1 molecules (sub-diffusion mechanism)**

**Gradient of membrane surface tension (the Marangoni effect)**

**In-plane surface tension: rearrangement of lipids and conformational changes of actin-cortex**

**Curvature induced-changes in the surface area lead to an increase in the out-of-plane surface tension**

**Figure 3**. A schematic presentation of the mechanisms that influence lateral migration of Piezo1 channels: Piezo1 channels diffuse laterally along the cell membrane in response to concentration gradients. This motion is hindered by resistive interactions with the cortical structure (i.e., subdiffusive behaviour). In the absence of additional driving forces, such diffusion tends to homogenize the spatial distribution of Piezo1 channels. However, membrane curvature introduces gradients in surface tension, giving rise to Marangoni-like flows that promote an inhomogeneous redistribution of Piezo1 and clustering around FAs. The membrane surface tension includes two contributions: in plane contributions account for conformational changes of actin cortex and lipid rearrangement, while out-of-plane contribution is induced by curvature-induced local change in the membrane surface area. Blue arrows represent various contributions to relevant physical parameters.

In further consideration, it is necessary to formulate the membrane surface tension contributions and the mass balance of Piezo1 channels based on the framework proposed by Pajic-Lijakovic et al. (2026) **[**21]. The membrane surface tension is formulated in **Box 1**:

**Box 1**. The membrane surface tension

The surface tension of the membrane can be expressed as [21]:

$$\gamma_m(r,\tau) = \gamma_m^{in-plane} + \gamma_m^{out-of-plane} \quad (1)$$

where $r$ is the radial coordinate near an FA, $\tau$ is the timescale of minutes, $\gamma_m^{in-plane}$ is the in-plane dilational component and $\gamma_m^{out-of-plane}$ is the curvature-dependent component of the membrane surface tension $\gamma_m$. The dilational component $\gamma_m^{in-plane}$ can be expressed in the form of a suitable constitutive model of dilational viscoelasticity $\Delta\gamma_m^{in-plane} - \frac{\Delta A}{A}$ (where $\gamma_m^{in-plane} = \gamma_c^{in-plane} + \gamma_{BL}^{in-plane}$, $\gamma_c^{in-plane}$ is the cortex contribution and $\gamma_{BL}^{in-plane}$ is the bilayer contribution, $\Delta\gamma_m^{in-plane} = \gamma_m^{in-plane}(r,\tau) - \gamma_m^{in-plane}(r,0)$, and $\frac{\Delta A}{A}$ is the change in the surface area). A change in the $\gamma_m^{in-plane}$ induces energy storage and dissipation. The out-of-plane, curvature-dependent component $\gamma_m^{out-of-plane}$ can be expressed as: $\gamma_m^{out-of-plane} = \alpha_c \frac{1}{2}\left|\vec{\nabla} h\right|^2$ (where $\alpha_c$ is the effective stiffness of the membrane and $h(r,\tau)$ is the membrane out-of-plane displacement).

The structural inhomogeneity present in the lipid bilayer and actin-cortex, along with the formation of curvature in the membrane, leads to a non-uniform distribution of surface tension across the membrane. The gradient in membrane surface tension drives the movement of Piezo1 molecules from regions of higher surface tension, located distantly from FAs, towards regions of lower surface tension found in the inwardly curved areas adjacent to FAs. This phenomenon is known as the Marangoni effect. This phenomenon already is already known in various soft matter systems, in which the gradient of surface tension is induced primarily by variation in temperature and the distribution of system constituents [73]. However, curvature also influences the surface tension [74]. The Marangoni effect has been discussed in the context of the migration of Piezo1 channels [21]. The mass balance equation of Piezo1 molecules, including their migration driven by the gradient of their concentration (diffusion mechanism) and the gradient of the membrane surface tension (Marangoni effect) is as shown in **Box 2**:

**Box 2**. The mass balance equation for Piezo1 molecules

Piezo1 molecules engage in hop diffusion across two domains of the actin cytoskeleton within a time frame of milliseconds [36]. Each successive hop results in damping effects on the migration of Piezo1 molecules [36]. The long-term diffusion reflects the cumulative impact of these successive hops, which effectively damps Piezo1 migration. The damping in Piezo1 lateral migration was characterized as sub-diffusion, which can be articulated through the application of fractional derivatives [17,75]. The mass balance equation for Piezo1 molecules can be expressed as [21]:

$$D_t^\alpha c_p(r,\tau) = -\vec{\nabla}_s \cdot \left(D_\alpha \vec{\nabla}_s c_p + \mu_p c_p \vec{\nabla}_s \gamma_m\right) \qquad (2)$$

where $c_p(r,\tau)$ is the surface number density of Piezo1 molecules, $\mu_p$ is their mobility driven by the gradient of the membrane surface tension, $D_\alpha$ is the anomalous diffusion coefficient in units of $\frac{\mathrm{m}^2}{\mathrm{s}^\alpha}$, $D_t^\alpha\left(\cdot\, c_p\right)$ is a fractional derivative, and $\alpha$ is its order which satisfies the condition $0 < \alpha < 1$, described as the damping coefficient. The fractional derivatives of a function $f(t)$ are equal to $D_t^\alpha\big(f(t)\big) = \frac{d^\alpha}{dt^\alpha} f(t)$. We used Caputo's definition of the fractional derivative as follows [75]: $D_t^\alpha\big(f(t)\big) = \frac{1}{\Gamma(1-\alpha)} \frac{d}{dt} \int_0^t \frac{f(t')}{(t-t')^\alpha} dt'$ (where $t$ is the independent variable time, and $\Gamma(1-\alpha)$ is the gamma function). If the order of the fractional derivative $\alpha = 0$, then $D_t^0\big(f(t)\big) = f(t)$. However, it can be shown that when $\alpha \to 1$, then $D_t^\alpha\big(f(t)\big) \to \frac{df(t)}{dt}$. The first term on the right-hand side of eq. 2 represents the diffusive flux, while the second represents the Marangoni flux.

Consequently, the formation of curvatures along FAs is a prerequisite for Piezo1 accumulation near FAs. Experimental evidence supports the tendency of Piezo1 channels to cluster near FAs in: (i) epithelial cells and fibroblasts in studies without LIUS [5] and (ii) mesenchymal-like cancer cells accompanied by epithelial cells and fibroblasts in investigations utilizing LIUS [6]. Nonetheless, the spatial distribution of Piezo1 molecules in these curved regions has yet to be investigated. It is known, however, that Piezo1 molecules grouped near FAs can be either active or inactive.

### 6.2 Causes of the curvature formation near FAs: the force balance equations

In-plane and out-of plane driving and resistive forces act on the ventral cytoskeleton under the FAs. They induce membrane fluctuations that generate strain equal to $\tilde{\boldsymbol{\varepsilon}}_{\boldsymbol{m}} = \frac{1}{2}\left(\vec{\boldsymbol{\nabla}}\,\vec{\boldsymbol{u}}_{\boldsymbol{m}} + \vec{\boldsymbol{\nabla}}\,\vec{\boldsymbol{u}}_{\boldsymbol{m}}^{\;\boldsymbol{T}}\right)$ (where $\vec{\boldsymbol{u}}_{\boldsymbol{m}}(\vec{\boldsymbol{u}}_{\boldsymbol{c}}(u_{mx}, u_{my}), h)$ is the displacement of the ventral cytoskeleton, $\vec{\boldsymbol{u}}_{\boldsymbol{c}}(u_{mx}, u_{my})$ is the in-plane displacement vector, and $h$ is the out-of-plane displacement). The lipid bilayer, actin cortex, and ventral cytoskeleton form a lamellar structure and consequently undergo parallel mechanical coupling [76]. It means that the strain within all layers is the same, while the total stress generated within the membrane represents the sum of stresses generated within each layer.

The in-plane force balance is expressed as: $\frac{1}{\Delta A} \xi_u \frac{d\vec{\boldsymbol{u}}_c}{d\tau} = \sum_i \vec{\boldsymbol{F}}_{\boldsymbol{di}}^{\boldsymbol{in-plane}} - \sum_i \vec{\boldsymbol{F}}_{\boldsymbol{ri}}^{\boldsymbol{in-plane}}$ (where $\xi_u$ is the in-plane drag coefficient, $\vec{\boldsymbol{F}}_{\boldsymbol{di}}^{\boldsymbol{in-plane}}$ are driving forces, $\vec{\boldsymbol{F}}_{\boldsymbol{ri}}^{\boldsymbol{in-plane}}$ are the resistive forces, and $\Delta A$ is the surface area of an FA). In-plane driving forces are the traction force $\vec{\boldsymbol{F}}_{\boldsymbol{t}}^{\boldsymbol{in-plane}}$ and ultrasound body force $\vec{\boldsymbol{F}}_{\boldsymbol{LIUS}}{}^{in-plane}$, while the resistive forces are: the viscoelastic force $\vec{\boldsymbol{F}}_{\boldsymbol{ve}}^{\boldsymbol{in-plane}}$, and surface tension forces $\vec{\boldsymbol{F}}_{\boldsymbol{st}}^{\boldsymbol{in-plane}}$.

The out-of-plane force balance is expressed as: $\frac{1}{\Delta A}\xi_h \frac{dh}{d\tau} = \sum_i F_{di}^{out-of-plane} - \sum_i F_{ri}^{out-of-plane}$ (where $\xi_h$ is the out-of-plane drag coefficient). These forces emerge as a result of the out-of-plane displacement $h$ and are consistently expressed in scalar form (where $F_{di}^{out-of-plane}$ are out-of-plane driving forces and $F_{ri}^{out-of-plane}$ are out-of-plane resistive forces). Out-of-plane driving forces are: the FA cluster–induced out-of-plane driving force and ultrasound force, while the out-of-plane resistive forces are composed of the viscoelastic force, traction force, lipid bilayer bending force, and surface tension force.

The surface tension force originates from the lipid bilayer–actin cortex composite and is given by the gradient of the membrane surface tension. The viscoelastic force, in turn, arises from the ventral stress fiber network and is described by the divergence of the mechanical stress, accounting for spatial heterogeneities in stress fiber organisation.

The out-of-plane driving forces, such as the ultrasound force and the FA cluster–induced driving force, are responsible for the generation of out-of-plane displacement $h$ in the ventral cytoskeleton. The out-of-plane resistive forces such as the bilayer bending force and the surface tension force vary between cell types, but these variations are not responsible for the observed variation in curvature properties. The main resistive force for curvature formation is the out-of-plane viscoelastic force of the ventral cytoskeleton. This force depends, not on the magnitude of mechanical stress accumulated within the ventral stress fiber network and its stiffness, but rather on the distribution of mechanical stress and is equal to the divergence of the mechanical stress [17]. As a result, the structurally inhomogeneous ventral cytoskeleton found in mesenchymal-like cancer cells experiences a sufficiently strong force to prevent curvature formation along FAs in the absence of LIUS. In contrast, the more uniform distribution of mechanical stress observed in epithelial cells and fibroblasts results in a reduction of the viscoelastic force in relation to the driving force induced by FA clusters, which consequently leads to the formation of curvatures along FAs even without LIUS.

Nevertheless, the application of ultrasound has the potential to diminish the viscoelastic force. There are two potential physical mechanisms. The first, which is more likely, suggests that LIUS creates a tilting effect in the network of ventral stress fibers. The second mechanism could be attributed to a mechanically-induced disintegration of the ventral cytoskeleton, resulting in energy dissipation and a decrease in the viscoelastic force. However, it is unlikely that LIUS alone can create extreme stress inhomogeneities sufficient to physically destabilize the cytoskeleton. We infer that it is the localized tilting is responsible for the development of irregular curvatures along FAs in cancer cells.

**6.2.1 The out-of-plane driving forces for the curvature formation**

An FA cluster–induced out-of-plane driving force of sufficient intensity can cause the formation of curvature near FAs in epithelial cells and fibroblasts, even without LIUS [41]. This force depends on: (i) the surface number density of integrin-ligand bonds $\rho_B$, (ii) the force per single bond $f_0$, and (iii) the tilting angle of integrin molecules $\vec{\boldsymbol{\varphi}}(\varphi_x, \varphi_y)$. The out-of-plane force density can be expressed as:

$$F_{FA}^{out-of-plane} = \rho_B(r,\tau) f_0(r,\tau)\cos(|\vec{\boldsymbol{\varphi}}(r,\tau)|) \quad (3)$$

where $|\vec{\boldsymbol{\varphi}}(r,\tau)| = \sqrt{\varphi_x^2 + \varphi_y^2}$. An increase in the tilting angle causes a decrease in the force $F_{FA}^{out-of-plane}$. Various membrane curvatures can be generated depending on the angle $\vec{\boldsymbol{\varphi}}$ when the out-of-plane viscoelastic force is relatively small**:** (i) for $\vec{\boldsymbol{\varphi}} = 0$, the curvature is relatively deep and symmetric, (ii) for $\vec{\boldsymbol{\varphi}} = \text{const}$ and $\vec{\boldsymbol{\varphi}} > 0$, the curvature is shallower and symmetric, and (iii) for $\vec{\boldsymbol{\varphi}}$ non-uniformly distributed along the FA, the curvature is irregular and consists of a few sub-curvatures. The non-uniform distribution of tilting angle can be caused by: (i) external factors such non-uniform stiffness of the substrate matrix and exposure of cells to LIUS. The $F_{FA}^{out-of-plane}$ is greatest in fibroblasts and lowest in migrating epithelial cells. However, despite the fact that cancer cells form larger FAs than epithelial cells, this force cannot generate curvature along FAs primarily due to the higher resistance force of the ventral cytoskeleton as shown in experiments without LIUS [5,6].

The other driving force, which acts both in-plane and out-of-plane, is the ultrasound force $\vec{\boldsymbol{F}}_{\boldsymbol{LIUS}}$**,** i.e., the acoustic body-force density, expressed as [77]:

$$\vec{\boldsymbol{F}}_{\boldsymbol{LIUS}}(r,\tau) = -\nabla\langle\tilde{\Pi}\rangle l_c \quad (4)$$

where $l_c$ is the thickness of the ventral cytoskeleton, $\tilde{\Pi}$ is the momentum flux equal to $\tilde{\Pi}(r,t,\tau) = p_{ac}\tilde{\boldsymbol{I}} + \rho_m \vec{\boldsymbol{v}}_{\boldsymbol{ac}} \otimes \vec{\boldsymbol{v}}_{\boldsymbol{ac}}$**,** $\rho_m$ is the density of the surrounding medium, $\tilde{\boldsymbol{I}}$ is the unit tensor, $p_{ac}$ is the acoustic pressure, $\vec{\boldsymbol{v}}_{\boldsymbol{ac}}$ is the velocity of the external medium around the cells, and $\tau$ is the timescale in minutes.

The acoustic pressure $p_{ac}(r,t,\tau) = p_0 + \epsilon p^{(1)} + \epsilon^2 p^{(2)}$, where $p_0$ is the hydrostatic pressure, $p^{(1)}$ is the short-time pressure fluctuations expressed as: $p^{(1)}(r,t) = \Delta p \sin(\omega t)$, $p^{(2)}(r,\tau)$ is the slow generation of pressure caused by inhomogeneity of the ventral cytoskeleton, which leads to the structural perturbation of ventral stress fibers, and $\epsilon$ is the small correction coefficient. The time-averaged pressure over the acoustic period of oscillation is equal to $\langle p_{ac}\rangle(r,\tau) = p_0 + \epsilon^2\langle p^{(2)}\rangle$, while the short-term contribution is $\langle p^{(1)}\rangle = 0$. The out-of-plane component of the ultrasound force $F_{LIUS}^{out-of-plane} = \vec{\boldsymbol{n}} \cdot \vec{\boldsymbol{F}}_{\boldsymbol{LIUS}}$ accompanied by the FA cluster–induced out-of-plane driving force causes the formation of curvatures along FAs, which leads to the accumulation of Piezo1 molecules in the regions around Fas in all of the cell types considered [6]. These driving forces act against the resistive forces. The main resistance to curvature formation comes from the ventral cytoskeleton and is expressed in terms of the viscoelastic force.

### 6.2.2 The viscoelastic force and tilting of ventral cytoskeleton in cancer cells

Although ultrasound has been shown to perturb cytoskeletal dynamics and FA–linked actin cytoskeleton in cells [37], direct evidence for specific tilting of ventral stress fibers under LIUS — particularly due to inhomogeneous mechanical force distributions in cancer cells — has not yet been demonstrated. Therefore, the proposed sequence in which LIUS reduces viscoelastic resistance and favors irregular inward curvature formation at focal adhesions reflects a physically motivated conceptual model that remains to be experimentally validated.

The viscoelastic force depends on the distribution of mechanical stress around the ventral stress fiber network and can be expressed as [21]:

$$\vec{\boldsymbol{F}}_{\boldsymbol{vis}} = l_c \nabla \widetilde{\boldsymbol{\sigma}}_{\boldsymbol{c}} \tag{5}$$

where $\widetilde{\boldsymbol{\sigma}}_{\boldsymbol{c}}$ is the mechanical stress accumulated within the ventral cytoskeleton, which depends on its viscoelasticity. The viscoelasticity of the ventral stress fiber network is discussed in the **Appendix.**

Tilting of the ventral stress fiber network by the angle $\vec{\boldsymbol{\theta}}(\theta_x, \theta_y)$ relative to the in-plane axis under LIUS rotates the stress tensor: $\widetilde{\boldsymbol{\sigma}}_{\boldsymbol{cR}}(\vec{\boldsymbol{\theta}}) = \widetilde{\boldsymbol{R}}(\vec{\boldsymbol{\theta}})\sigma_{czz}\widetilde{\boldsymbol{R}}^{\boldsymbol{T}}(\vec{\boldsymbol{\theta}})$ (where $\widetilde{\boldsymbol{R}}(\vec{\boldsymbol{\theta}})$ is the rotation tensor, $\widetilde{\boldsymbol{\sigma}}_{\boldsymbol{cR}}$ is the stress tensor, which accounts for local rotation caused by the tilting of ventral cytoskeleton, and $\sigma_{czz}$ is the *z*-component of normal stress before rotation). The out-of-plane component of the viscoelastic force is equal to $F^n_{vis} = l_c \vec{\boldsymbol{n}}(\vec{\boldsymbol{\theta}}) \cdot \nabla \left(\widetilde{\boldsymbol{R}}(\vec{\boldsymbol{\theta}})\sigma_{czz}\widetilde{\boldsymbol{R}}^{\boldsymbol{T}}(\vec{\boldsymbol{\theta}})\right)$ (where $\vec{\boldsymbol{n}}$ is the local unit normal to the membrane). In turn, the out-of-plane viscoelastic force can be formulated here as:

$$F^n_{vis} \approx F^n_{vis0} \cos\left(\left|\vec{\boldsymbol{\theta}}\right|\right) \tag{6}$$

where $\left|\vec{\boldsymbol{\theta}}\right| = \sqrt{\theta_x^2 + \theta_y^2}$, and the out-of-plane viscoelastic force without tilting is expressed as $F^n_{vis0} = \vec{\boldsymbol{n}} \cdot l_c \nabla \widetilde{\boldsymbol{\sigma}}_{\boldsymbol{c}}$**.** An increase in the angle $\vec{\boldsymbol{\theta}}$ results in a decrease in the force $F^n_{vis}$. In this case, the out-of-plane ultrasound force accompanied by the FA cluster–induced driving force can generate curvatures along FAs in mesenchymal-like cancer cells. Tilting-induced curvatures are irregular. The tilting angle $\vec{\boldsymbol{\theta}}$ has a feedback impact on the tilting angle of integrin $\vec{\boldsymbol{\varphi}}$**.** Spatio-temporal change of the tilting vector under LIUS can be expressed here as:

$$\eta_\theta \frac{d\vec{\boldsymbol{\theta}}}{d\tau} = D_\theta \nabla^2 \vec{\boldsymbol{\theta}} - \frac{1}{\tau_\theta}\vec{\boldsymbol{\theta}} + \sum_{\boldsymbol{i}} \vec{\boldsymbol{M}}_{\boldsymbol{di}} - \sum_{\boldsymbol{j}} \vec{\boldsymbol{M}}_{\boldsymbol{rj}} \tag{7}$$

where $\eta_\theta$ is the rotational viscosity, $D_\theta$ is the diffusion coefficient of tilt along the ventral fiber network, $\tau_\theta$ is the relaxation time of the ventral cytoskeleton, $\vec{\boldsymbol{M}}_{\boldsymbol{di}} = \vec{\boldsymbol{r}} x \vec{\boldsymbol{F}}_{\boldsymbol{di}}$ is the *i*-th torque made by the driving force $\vec{\boldsymbol{F}}_{\boldsymbol{di}}$**,** and $\vec{\boldsymbol{M}}_{\boldsymbol{rj}} = \vec{\boldsymbol{r}} x \vec{\boldsymbol{F}}_{\boldsymbol{rj}}$ is the *j*-th torque of the resistive force $\vec{\boldsymbol{F}}_{\boldsymbol{rj}}$**.** Besides these forces, the traction force can be either a driving or a resistive force.

### 6.2.3 Traction force

The traction force depends on the structural homogeneity, viscoelasticity, and stiffness distribution of a substrate matrix. The relationship between the traction force and the displacement of the substrate can be presented in the form of the Zener constitutive model [78]:

$$\tau_F \frac{d\vec{\boldsymbol{F}}_{\boldsymbol{t}}}{d\tau} + \vec{\boldsymbol{F}}_{\boldsymbol{t}} = E_s(r)\vec{\boldsymbol{u}}_{\boldsymbol{s}} + \eta_s(r)\frac{d\vec{\boldsymbol{\varepsilon}}_s}{d\tau} \tag{8}$$

where $\vec{\boldsymbol{F}}_{\boldsymbol{t}}(r,\tau)$ is the traction force per unit area of matrix, $\vec{\boldsymbol{\varepsilon}}_{\boldsymbol{s}}$ is the dimensionless displacement field $\vec{\boldsymbol{\varepsilon}}_{\boldsymbol{s}} = \frac{\vec{\boldsymbol{u}}_s}{\langle \vec{\boldsymbol{u}}_s \rangle}$, $\vec{\boldsymbol{u}}_{\boldsymbol{s}}$ is the matrix displacement field, $\langle \vec{\boldsymbol{u}}_{\boldsymbol{s}} \rangle$ is the averaged matrix displacement field, $\tau_F$ is the relaxation time of the traction force under constant displacement, $E_s(r)$ is the elastic modulus of the matrix, and $\eta_s(r)$ is its viscosity. This force pulls the ventral cytoskeleton horizontally toward the FA. The magnitude of the traction force depends on the stiffness of the substrate matrix. Cancer cells lack the ability to perceive matrix rigidity, leading to significantly dynamic focal adhesions (FAs) [79]. The in-plane traction force $\vec{\boldsymbol{F}}_{\boldsymbol{t}}^{\,in-plane} = \left(\vec{\boldsymbol{t}} \cdot \vec{\boldsymbol{F}}_{\boldsymbol{t}}\right) \cdot \vec{\boldsymbol{t}}$ is a driving force for the remodelling of the ventral cytoskeleton, while the out-of-plane traction force $F_t^n = \vec{\boldsymbol{n}} \cdot \vec{\boldsymbol{F}}_{\boldsymbol{t}}$ resists this remodelling (where $\vec{\boldsymbol{t}}$ is the tangential unit vector). Inhomogeneity of the ventral cytoskeleton in cancer cells leads to inhomogeneity of the transmitted cell tractions [80]. Other resistive forces are the surface tension force and the lipid bilayer bending force.

### 6.2.4 Other resistive forces

The surface tension force resists increase in the surface area of the membrane. The in-plane surface tension force represents a consequence of the inhomogeneous distribution of the in-plane surface tension and can be expressed as: $\vec{\boldsymbol{F}}_{\boldsymbol{st}}^{\boldsymbol{in-plane}} = \vec{\boldsymbol{\nabla}} \gamma_m^{in-plane}$, while the out-of-plane surface tension force $F_{st}^{out-of-plane} = \vec{\boldsymbol{\nabla}} \cdot \left(\gamma_m^{out-of-plane} \vec{\boldsymbol{\nabla}} h\right)$ depends on the characteristic of the curvature [21]. Consequently, the surface tension force can resist the formation of irregular curvatures along FAs. The other resistive force is the out-of-plane lipid bilayer bending force $F_{bend} = -\frac{\delta E_{bend}}{\delta h}$ (where $E_{bend}$ is the bending energy of the lipid bilayer $E_{bend} = \frac{1}{2} E_b \left(\frac{\partial^2}{\partial x^2} h + \frac{\partial^2}{\partial y^2} h - 2C_0\right)^2$ and $E_b$ is the bending modulus, which depends on the concentration of cholesterol in curved regions [21,36]. A higher concentration of cholesterol increases the bending modulus and consequently resists bending of the bilayer. The modulation of Piezo1 activity by Irregular curvature along FAs will be discussed in the context of established physical parameters.

## 7. The relationship between physical parameters

Clustering of the Piezo1 channels around focal FAs in cells when exposed to low-frequency low-intensity ultrasound (LIUS) is likely a cell-state dependent phenomenon rather than a universal one. Thus, the Piezo1 channel clustering around FAs in all cell types under the influence of LIUS is not fully supported by the current research as, for example, shown for red blood cells [36]. Nevertheless, the clustering of Piezo1 molecules around FAs marks the formation of curvatures within these structures. As our theoretical model suggests the curvature formation is driven primarily by the interaction between the out-of-plane driving force induced by integrin clustering, and the viscoelastic force which reflects the resistive effects of the ventral cytoskeleton. These resistive effects are influenced by the inhomogeneous distribution of mechanical stress. Additionally, the tilting of the ventral cytoskeleton has the potential to diminish its resistance. In cancer cells, the ventral cytoskeleton is characterized by an inhomogeneous distribution of stress fibers, in contrast to the more uniform distribution observed in fibroblasts and epithelial cells. As a result, the viscoelastic force is pronounced in cancer cells, which can counteract the formation of curvatures along FAs in response to the driving force. Furthermore, Piezo1 molecules are uniformly distributed along the membranes of cancer cells, rather than being concentrated at FAs [5,21]. **Figure 3** provides a schematic representation of the mechanism through which application of LIUS leads to calcium-induced apoptosis of cancer cells:

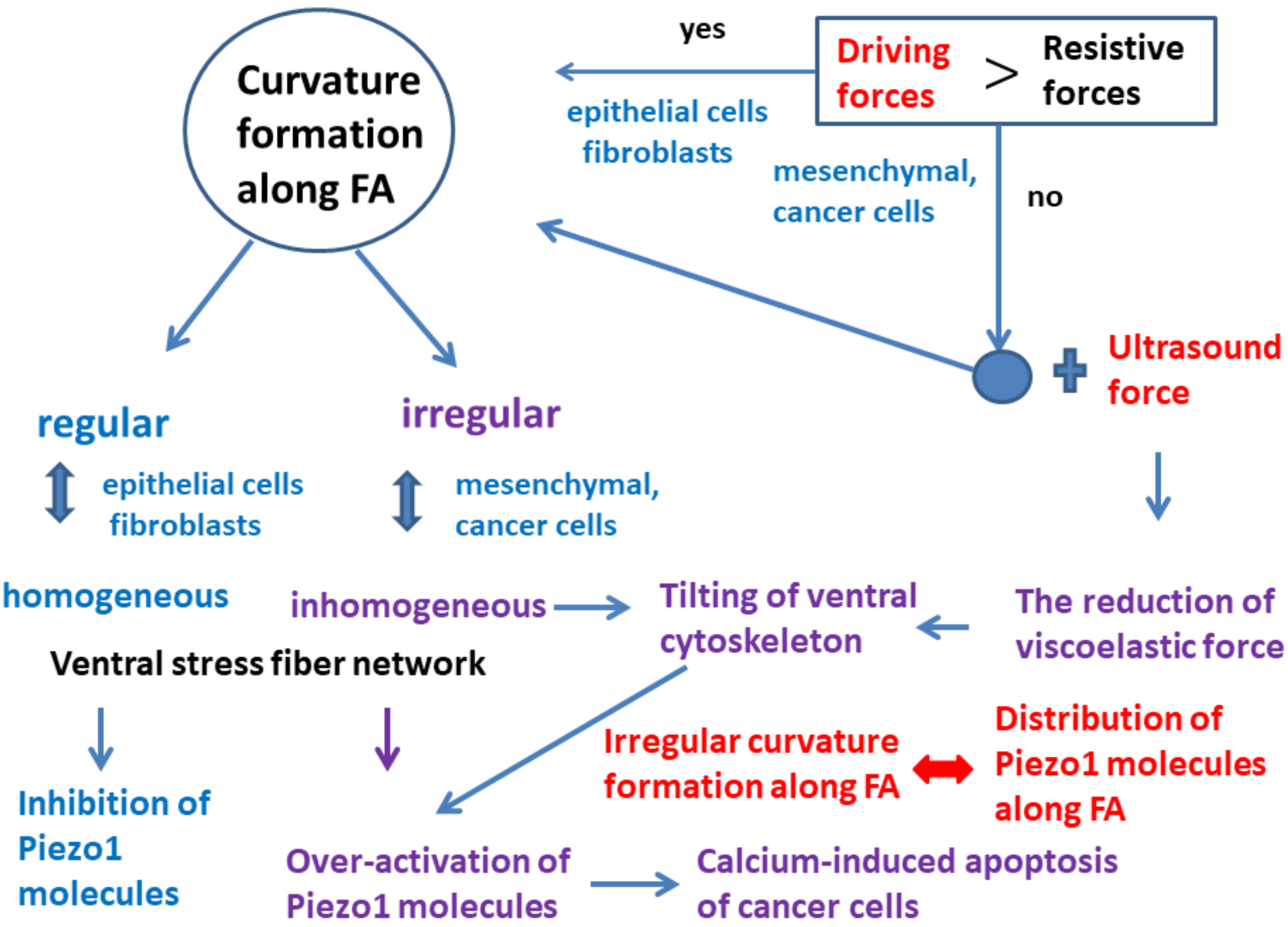

**Figure 4**. The scenario through which the application of LIUS leads to the calcium-induced apoptosis of cancer cells

The primary factors influencing curvature formation are the density and distribution of the strength of integrin-ligand bonds, along with the tilting of integrin molecules. Their tilting within an FA, caused by stiffness inhomogeneity and topological defects of the substrate matrices, can reduce the $F_{FA}^{out-of-plane}$ force. Although this force induced by FA clusters is minimal in epithelial cells, it remains sufficiently strong to promote the development of curvatures along FAs, counteracting the viscoelastic force.

Nevertheless, when the ultrasound force is applied in conjunction with the out-of-plane driving force induced by the integrin cluster, curvatures develop along the focal adhesions in mesenchymal-like cancer cells. This phenomenon can be attributed to two primary factors. Firstly, the ultrasound force facilitates the formation of curvatures, and secondly, it can induce a tilting of the ventral cytoskeleton, which further reduces the resistive viscoelastic force. The tilting of the ventral cytoskeletons of cancer cells subjected to LIUS arises from their structural inhomogeneity.

As a result, we propose that Piezo1 molecules are clustered around FAs possibly in all cell types when exposed to low frequency LIUS [6]. In cancer cells, these molecules exhibit heightened activity, resulting in calcium-induced apoptosis; however, in epithelial cells and fibroblasts, the Piezo1 channels demonstrate significantly reduced activity. The activity of Piezo1 channels is influenced by their distribution within curved areas. Furthermore, the arrangement of Piezo1 molecules is contingent upon the dimensions and regularity of these curved regions.

In regions of larger regular curvature, Piezo1 molecules exhibit a denser packing and self-inhibited opening in relation to one another similar to what has been shown for MscL channels [45]. Sufficient membrane cholesterol supports Piezo1 cluster organization and coordinated activity, whereas cholesterol depletion increases the gating force and alters channel kinetics, as demonstrated in multiple studies [19,71,72]. Conversely, irregular curvatures, which comprise several smaller sub-curvatures, can create niches for Piezo1 molecules, effectively separating them from one another. In such instances, Piezo1 molecules maintain their activity. The tilting of the ventral cytoskeleton under low-intensity ultrasound (LIUS) can affect the irregularity of the curvatures along focal adhesions (FAs).

The irregularity of the curvatures is a primary factor in the activation of Piezo1 molecules in cancer cells under LIUS: it facilitates a significant influx of calcium into these cells, ultimately leading to their apoptosis.

**8. Predictive consequences of curvature-mediated Piezo1 recruitment under LIUS**

The proposed framework yields experimentally testable predictions regarding the differential response of epithelial, fibroblast, and mesenchymal-like cancer cells to LIUS. In particular, the model predicts that LIUS-induced perturbation of the ventral cytoskeleton can alter local membrane curvature in FA–associated regions through changes in stress fiber organization and viscoelastic resistance. Because Piezo1 recruitment is expected to depend on local curvature and membrane mechanics, mesenchymal-like cancer cells with spatially heterogeneous ventral cytoskeletal organization are predicted to exhibit enhanced local Piezo1 accumulation and mechanosensitive activation relative to epithelial cells and fibroblasts. A key consequence of this framework is that curvature-mediated redistribution of Piezo1 may shift channels from mechanically independent to mechanically coupled regimes, thereby amplifying downstream calcium signalling responses. The mechanical coupling between Piezo1 footprints can lead to enhanced or reduced channel activation, depending on the spatial regularity of membrane curvature in the vicinity of focal adhesions. Consequently, two key physical parameters can be identified as primary determinants of Piezo1 activation under LIUS: the local packing density of Piezo1 channels within clusters around FA, and the spatial regularity of membrane curvature in their vicinity.

The average surface density of Piezo1 channels in the plasma membrane is relatively low ($\sim 2 \ \frac{\text{channels}}{\mu\text{m}^2}$), but local clustering in functional regions such as FAs or neurite-like protrusions can increase the effective density by an order of magnitude [5,31,42]. Within the present framework, such spatial redistribution is critical because it determines whether Piezo1 footprints operate in an independent or mechanically coupled regime.

A key prediction of the model is the existence of a critical surface density ($n_c$), above which curvature-mediated coupling between Piezo1 footprints becomes significant. This threshold is given by $n_c = \frac{1}{R_f^2 \pi}$, $R_f$ is the average radius of footprint equal to $R_f \sim 3\lambda$ and λ is decay length [43]. The decay length can be estimated as $\lambda = \sqrt{\frac{E_{beff}}{\gamma_m}}$,and $E_{beff}$ is the effective bending modulus of the lipid bilayer equal to: $E_{b0} \approx 20 k_B T$ (where $k_B$ is Boltzmann's constant and $T$ is temperature) [81]. The effective bending modulus for rigid inclusions, such as Piezo1 channels, can be expressed as: $E_{beff} = E_{b0}\left(1 + \phi_p\right)$ (where $\phi_p$ is the surface packing density of Piezo1 channels within cluster). The surface tension of the cell membrane, which includes the contributions of the lipid bilayer and actin-cortex, is in the range of $\gamma_m \sim 0.1 - 1 \ \frac{\text{mN}}{\text{m}}$ [82]. For a surface packing density of Piezo1 channels equal to $n_c \sim 50 \ \frac{\text{channels}}{\mu\text{m}^2}$, the surface fraction is $\phi_p \sim 0.016$. The calculated footprint radius based on the procedure proposed by Haselwandter and MacKinnon [43] is $\sim 50 - 60$ nm. Consequently, the footprint diameter corresponds to the order of magnitude of the average size of the cortex meso-domain.

When the packing density of Piezo1 channels is $n < n_c$, Piezo1 channels preferentially adopt outwardly curved membrane footprints, corresponding to an energetically favorable minimum of the membrane–protein system [18]. The outward membrane footprint of individual Piezo1 channels is stabilized in curved membrane regions and is resistant to perturbation [18]. However, when local density exceeds the critical threshold and curvature fields become sufficiently regular and inward, mechanical coupling between neighboring footprints can destabilize this outward configuration and promote transitions toward inward footprints. This collective effect is strongly enhanced in geometrically regular curvature fields, where footprint interactions are coherent [42], whereas curvature irregularity can disrupt coupling and preserves the outward, active state.

Consequently, curvature regularity emerges as a second control parameter, in addition to local density, governing Piezo1 activity. Regular curvature promotes collective footprint coupling and reduced activity, while irregular curvature suppresses coupling and maintains channel activation. Within this framework, cancer cells and epithelial/fibroblast cells are distinguished not only by differences in clustering density, but also by differences in curvature regularity under external mechanical stimulation.

Finally, this provides a direct mechanistic link to LIUS: ultrasound amplitude and wavelength determine both curvature magnitude and spatial regularity, thereby controlling whether the system operates above or below the coupling threshold and whether Piezo1 channels remain active or become collectively inhibited. This establishes a predictive mechanical criterion for Piezo1 activation based on the interplay between channel density, footprint size, and curvature geometry.

From an experimental perspective, the above predictions can be directly tested by quantifying how low-frequency ultrasound modulates curvature statistics at FAs and how these changes correlate with Piezo1 redistribution. In particular, several key model variables can be mapped onto experimentally measurable quantities. The tilting of ventral stress fibers can be quantified through the local angular distribution of cytoskeletal filaments, providing a direct measure of force-transmission reorganization under LIUS. This can be achieved using high-resolution fluorescence imaging techniques, including confocal microscopy, total internal reflection fluorescence (TIRF) microscopy, or super-resolution approaches such as STED or structured illumination microscopy (SIM), combined with quantitative image analysis of filament orientation [83-85]. The characterization of structural inhomogeneity (disorder) of cytoskeleton in mesenchymal-like cancer cells by the procedure proposed by Damania et al. [12] using PWS microscopy has been expanded into broader clinical and biophysical applications [86]. In parallel, FA–associated curvature and dynamics can be assessed using membrane reporters and live-cell imaging, while Piezo1 redistribution can be quantified through fluorescently tagged constructs or immunofluorescence-based localization analysis [5]. Together, these approaches enable a direct experimental validation of the predicted coupling between LIUS-induced cytoskeletal reorganization, curvature formation, and Piezo1 recruitment.

The resulting curvature field around FAs can be characterized experimentally through spatial curvature maps, where curvature heterogeneity can be quantified using the geometric variance $\nu_C(r,\tau)$ expressed as: $\nu_C(r,\tau) \sim \langle (\vec{\nabla}^2 C_{FA})^2 \rangle - \langle \vec{\nabla}^2 C_{FA} \rangle^2$ (where $C_{FA}$ is the curvature around FA). LIUS amplitude and spatial characteristics are expected to control both the magnitude and spatial organization of curvature fields generated at focal adhesion sites. These curvature fields can be characterized through the variance of local curvature and its spatial correlation length, providing quantitative measures of curvature heterogeneity versus coherence. Within this framework, different mechanical cell states are expected to exhibit distinct curvature responses under identical LIUS exposure: cells with heterogeneous ventral cytoskeletal organization are predicted to generate more spatially variable curvature fields, whereas cells with more homogeneous force transmission are expected to exhibit more spatially coherent curvature responses. Transitions between weak and strong Piezo1 coupling regimes can then be identified by measuring the coupled evolution of curvature regularity, Piezo1 surface fraction, and channel clustering statistics under controlled ultrasound stimulation.

A key prediction of the model is that, at fixed channel density, increasing curvature heterogeneity suppresses cooperative Piezo1 coupling and maintains channel activity, whereas increasing curvature coherence enhances collective coupling effects and promotes channel inhibition. This establishes a predictive mechanical criterion for Piezo1 activation based on the interplay between channel density, footprint size, curvature geometry, and externally controlled LIUS parameters.

### 8.1 Mechanistic strategies for improving the selectivity of LIUS-based cancer treatment

An important direction for improving the selectivity of LIUS-based cancer treatment is the amplification of pre-existing mechanical differences between normal and cancer cells. In principle, susceptibility to LIUS may be enhanced through interventions that modulate cellular mechanics, mechanotransduction, or cell–matrix interactions. Mechanically, altered pre-stress or cytoskeletal organization could increase the sensitivity of mesenchymal-like cancer cells to ultrasound-induced deformation. Pharmacologically, modulation of actomyosin contractility, calcium-dependent signaling, focal adhesion turnover, or mechanosensitive pathways may alter the dynamic mechanical response of cancer cells. Matrix-based interventions that modify extracellular matrix stiffness, confinement, or adhesion strength may further bias force transmission and cytoskeletal remodeling. Because cancer cells frequently exhibit altered contractility, FA mechanics, and mechanically heterogeneous intracellular organization, such perturbations could potentially increase LIUS susceptibility while preserving the relative resistance of normal epithelial cells. These strategies are presented as experimentally testable directions for future investigation rather than as established therapeutic approaches.

## 9. Conclusion

This theoretical analysis identifies curvature geometry along focal adhesions as a key physical factor governing the selective activation of Piezo1 channels in cancer cells exposed to low-frequency ultrasound. The geometry of the curvature depends on the structural organization of the ventral cytoskeleton. Epithelial cells and fibroblasts have more homogeneous ventral cytoskeleton than cancer cells.

Fibroblasts and cancer cells typically exhibit an anisotropic ventral cytoskeletal organization, whereas epithelial cells generally display a more isotropic organization. Both fibroblasts and mesenchymal-like cancer cells possess aligned stress fiber networks coupled to FAs; however, in cancer cells the ventral cytoskeletal architecture is more spatially heterogeneous, leading to a non-uniform distribution of mechanical stress. In our continuum description, the viscoelastic resistive force is identified with the divergence of the surface stress tensor corresponding to the local force density within the ventral cytoskeleton. Spatial heterogeneity of the stress field can therefore suppress coherent curvature formation at FAs, despite local out-of-plane driving forces associated with FA clustering. However, LIUS introduces an additional out-of-plane driving force and can simultaneously induce tilting of the ventral cytoskeleton, by reducing the effective viscoelastic resistance and enabling curvature formation along FAs in these cells. As a result, Piezo1 channels, initially uniformly distributed along the membrane, are recruited into these newly formed curved regions.

A critical outcome of this process can be the emergence of highly irregular curvature patterns in cancer cells, consisting of multiple small sub-curvatures. These irregular topographies can prevent dense Piezo1 packing and self-inhibition, preserve channel activity, and reduce local cholesterol depletion—factors that collectively sustain Piezo1 opening under LIUS. The resultant strong calcium influx drives apoptosis specifically in cancer cells.

In contrast, epithelial cells and fibroblasts possess more homogeneous ventral cytoskeletal organization and more regular integrin–ligand bond distributions. These features promote the formation of larger, smoother curvatures along FAs, where Piezo1 channels pack more densely, experience self-inhibition, and show reduced activity even under LIUS. Consequently, these healthy cells maintain their viability and proliferative capacity.

In summary, the framework of our study suggests that LIUS selectivity arises not from biochemical differences but from intrinsic mechanical vulnerabilities of cancer cells—particularly their propensity to form irregular FA-associated curvatures that permit sustained Piezo1 activation. These findings highlight curvature-mediated mechanotransduction as a potential physical mechanism for targeting malignant cells and provide a foundation for designing ultrasound-based therapies that exploit cytoskeletal heterogeneity in cancer.

**Conflict of interest**: The authors report no conflict of interest.

**Funding**: This work was supported in part by the Engineering and Physical Sciences Research Council, United Kingdom (grant number EP/X004597/1), by the Ministry of Science, Technological Development and Innovation of the Republic of Serbia (Contract No. 451-03-34/2026-03/ 200135) and National Health and Medical Research Council of Australia (Investigator Grant L3 2034293).

**Appendix: The viscoelasticity of the ventral stress fiber network: the constitutive model**

A constitutive stress-strain model aimed at characterizing the viscoelastic properties of the ventral cytoskeleton can be developed based on micro-rheological studies. Micro-rheological investigations, represented through the relationship of storage and loss moduli against angular velocity, have been conducted for: (i) active actin networks in vitro [87-89] and (ii) membranes from various living cells [90-92]. The storage modulus represents a measure of the accumulated elastic and contractile energy, while the loss modulus represents a measure of energy dissipation. The experimental findings concerning live cells encompass the overall response of the dorsal region of the cell membrane, which integrates multiple contributions from the lipid bilayer, actin cortex, and cytoskeleton. Isolating the specific contribution of the cytoskeleton from these data presents a significant challenge. While the viscoelasticity of individual stress fibers corresponds to the Kelvin-Voigt model [62], the viscoelasticity of the ventral cytoskeleton should account for damping effects caused by inter-fiber interactions and has been described by including fractional derivatives in the form of the fractional Kelvin-Voigt model [93]. This model has described a wide range of *in vitro* and *in vivo* active actin networks. The fractional Kelvin-Voigt model, developed by Pajic-Lijakovic and Milivojevic, is articulated as [93]:

$$\tilde{\boldsymbol{\sigma}}_{\boldsymbol{c}}(r,\tau) = G_{sC}(r)\tilde{\boldsymbol{\varepsilon}}_{\boldsymbol{m}} + \eta_{\alpha}(r) D_t^{\alpha}\, \tilde{\boldsymbol{\varepsilon}}_{\boldsymbol{m}} + \eta_C(r)\dot{\tilde{\boldsymbol{\varepsilon}}}_{\boldsymbol{m}} \quad (1A)$$

where $\tau$ is the timescale in minutes, $\tilde{\boldsymbol{\sigma}}_{\boldsymbol{c}}(r,\tau)$ is the mechanical stress of the cytoskeleton, $G_{sC}(r)$ is the elastic modulus of the cytoskeleton, which accounts for elastic and actomyosin contractile energy, $\eta_{\alpha}(r)$ is the effective modulus, $\eta_C(r)$ is the viscosity of the cytoskeleton, $\tilde{\boldsymbol{\varepsilon}}_{\boldsymbol{m}}(r,\tau)$ is the strain of the cytoskeleton caused by membrane fluctuations, $\dot{\tilde{\boldsymbol{\varepsilon}}}_{\boldsymbol{m}}(r,\tau)$ is the strain rate equal to $\dot{\tilde{\boldsymbol{\varepsilon}}}_{\boldsymbol{m}} = \frac{d\tilde{\varepsilon}_m}{d\tau}$**,** and $D_t^{\alpha}\,(\tilde{\boldsymbol{\varepsilon}}_{\boldsymbol{m}})$ is a fractional derivative; $\alpha(r)$ is the order of the fractional derivative satisfying the condition $0 < \alpha < 1$, described as the damping coefficient [75]. For an inhomogeneous cytoskeleton, the model parameters: $G_{sC}$, $\eta_{\alpha}$, $\eta_C$, and $\alpha$ vary in space. The first term on the right-hand side of eq. 1A describes regular energy storage, while the third term quantifies regular energy dissipation within the cytoskeleton. The second term quantifies the anomalous nature of energy storage and dissipation within the cytoskeleton. Under LIUS, the local oscillatory strain is non-uniform, because stiffer domains deform less than softer regions, which can perturb the structural organisation of the cytoskeleton**.**